\documentclass[reprint, showpacs, nofootinbib, amsmath, amssymb,showkeys, aps, prl, superscriptaddress, secnum, nobibnotes
]{revtex4-2}
\usepackage{graphicx} 
\usepackage{dcolumn} 
\usepackage{bm} 
\usepackage{siunitx}
\usepackage{upgreek}
\usepackage{color}
\usepackage{hyperref} 
\usepackage{multirow}
\usepackage{xcolor}
\usepackage{float}
\usepackage{booktabs} 
\usepackage{nicefrac}
\hypersetup{hidelinks}

\begin{document}

\renewcommand{\thesection}{\Roman{section}} 
\setcounter{secnumdepth}{2}

\title{Design of a Fast Reactive Tuner for 1.3\,GHz TESLA cavities at MESA}
\author{Ricardo Monroy-Villa}
\email{rmonroyv@uni-mainz.de}
\affiliation{Institute for Nuclear Physics, Johannes Gutenberg-Universität Mainz, Mainz, Germany} 
\author{Ilan Ben-Zvi}
\email{Ilan.Ben-Zvi@StonyBrook.edu}
\affiliation{Physics and Astronomy Department, Stony Brook University, New York, USA}
\author{Florian Hug}
\email{flohug@uni-mainz.de}
\affiliation{Institute for Nuclear Physics, Johannes Gutenberg-Universität Mainz, Mainz, Germany} 
\author{Timo Stengler}
\email{stengler@kph.uni-mainz.de}
\affiliation{Institute for Nuclear Physics, Johannes Gutenberg-Universität Mainz, Mainz, Germany}

\date{\today} 
             
\begin{abstract}
This work presents a state-of-the-art design of a Ferroelectric Fast Reactive  Tuner (FE-FRT), capable of modulating high reactive power in TESLA type cavities on a microsecond time scale. The Mainz Energy-Recovering Superconducting Accelerator employs superconducting radio frequency cavities operating at 1.3 GHz, achieving quality factors on the order of $10^{10}$. However, detuning of $\pm$25\,Hz induced by microphonics have led to the use of strong coupling for the fundamental power coupler, requiring high-power amplifiers, orders of magnitude above the intrinsic dissipation. Current solutions to mitigate microphonics rely on piezoelectric tuners, which are not fast enough for the spectral range of the microphonics. A novel alternative is the FE-FRT, a technology made possible by the development of low-loss ferroelectric materials, which offer sub-microsecond response times.  Analytical results are provided along with their validation through finite-element simulations. The FE-FRT is expected to handle substantial reactive power while offering a tuning range of 50\,Hz in these type of cavities, resulting in a reduction in peak forward RF power by about an order of magnitude. 
\end{abstract}

\maketitle

\section{\label{sec:intro}Introduction}
Particle accelerators play a key role in fundamental research. In order to reach high energies,  electrons are often accelerated by superconducing radiofrequency (SRF) cavities, providing accelerating fields of tens of MV/m at quality factors of the order of a few times $10^{10}$. However, a common issue is microphonics, which causes a detuning of the resonant frequency, depending on the environment, of up to $\pm$25\,Hz.

The Mainz Energy-Recovering Superconducting Accelerator (MESA) is designed for medium-energy electron beams, focusing on high beam intensity and precision. The peak detuning of SRF cavities at MESA has been measured to be $\Delta f_{\mu}$=$\pm$25\,Hz, requiring to set the external quality factor of the fundamental power coupler to about $10^{7}$ to correct for microphonics \cite{mesamicro}. This common technique is applied in several facilities and has the setback of requiring high power RF sources as most of the forward power gets reflected. In an effort to mitigate this detuning, stepping motor and piezoelectric actuator tuners are employed. However, the response time of such tuners is not sufficiently fast to correct for higher frequency microphonics, which prevents increasing the external quality factor of the fundamental power coupler and thus limits the reduction of reflected power.

The development of low loss ferroelectric (FE) materials enables a new class of tuners, called Ferroelectric Fast Reactive Tuners (FE-FRT) which exhibit a sufficiently large tuning range as well as extremely fast response time, far beyond any microphonic perturbation. The MESA team conducts research towards the implementation of FE-FRT tuners to reduce the RF power consumption of MESA by more than an order of magnitude.

These FE materials are based on BaTiO$_3$/SrTiO$_3$-Mg ceramics, having a loss tangent, taken here conservatively as $\mathrm{tan}\left(\delta\right)\approx\delta$=2.39$\cdot10^{-3}$ at 1.3\,GHz, thermal conductivity $K$=7.02\,W/m and breakdown electric field of 20\,MV/m. When applying an electric field through the FE up to 8\,MV/m at 50\,°C, its relative permittivity changes from approximately $\epsilon_2=$129.60 to $\epsilon_1=$96.41, which are two ends states called state 2 or unbiased and state 1 or biased, respectively. This voltage-controlled permittivity is used to modulate the reactance of the tuner connected to the cavity, and thus change the cavity's frequency. The FE have been developed and thoroughly investigated by Euclid Techlabs Inc. \cite{euclid,aipeuclid} and used for the development of FE-FRTs along with CERN \cite{erl19,cern} and others \cite{hzb,hzbmag}. The response time capability of the FE in bulk form has been measured at $<$30\,ns. The first demonstration of the FE-FRT coupled to a 400\,MHz cavity was carried out at CERN in 2019 \cite{SRF19}, showing a frequency shift speed on the order of 600\,ns. 

In this work, we present the design of an FE-FRT to counteract the detuning effect originating from microphonics in 1.3\,GHz SRF 9-cell cavities of TESLA/XFEL type.  

\section{Analytical}
\begin{figure}[!htbp]
	\centering
		\includegraphics[width=1.00\columnwidth]{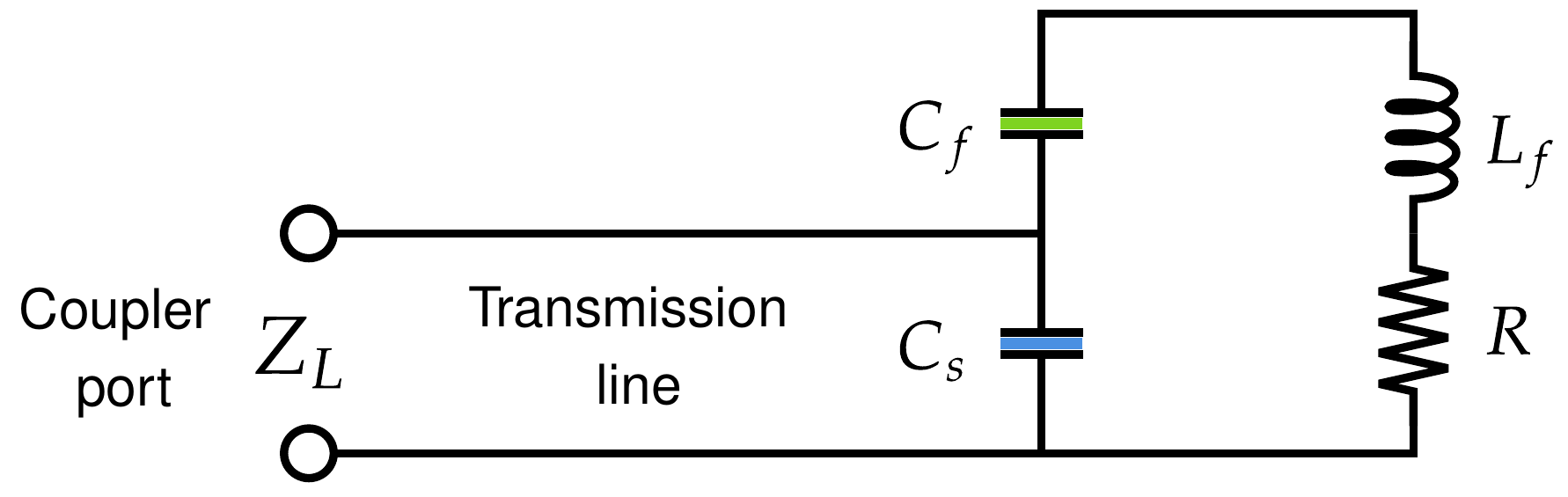}
	\caption{FE-FRT lumped-element circuit model. This figure is adapted from Figure\,1 in \cite{optim}.}    	
	\label{fig:FEFRTc}
\end{figure}

The tuner's layout and choice of parameters follows the procedure described in \cite{optim}. The equivalent circuit of the tuner is as shown in Figure\,\ref{fig:FEFRTc}. The complete tuner comprises a resonant circuit, which includes an inductance and two capacitors in series: The ferroelectric capacitor for tuning the frequency of the resonant circuit, and a series capacitor, used for coupling the resonator to a transmission line leading to the cavity's port. The length of the transmission line is chosen to be a quarter-wavelength (or odd multiple of quarter wavelengths).  The analytic expressions were evaluated using a code written in Maple \cite{maple2024}, then simulated in CST \cite{CST} code, including the tuner, a 1.3\,GHz 9-cell cavity and all associated coupling ports.
The analytical tuner parameters were optimized for cavity parameters of frequency $f_0$, stored cavity energy $U$ and tuning range $\Delta f$ as given in Table\,\ref{tab:table1}. 
Based on the comparison of the $FoM$ for various numbers of wafers \cite{optim} and the simplicity of the tuner assembly, a two-wafer annulus-shaped \cite{prab} ferroelectric capacitor was selected.

Then choices were made for the ferroelectric capacitor's gaps $g=0.5$\,mm and annulus width $w=0.5$\,mm. Given these choices, thermal considerations of the maximum allowed temperature rise in the ferroelectric material determined the following three tuner parameters:  the ratio of ferroelectric wafer area to thickness $A_{opt}$, the outer radius of the wafers $a_r$ and the capacitance of the ferroelectric capacitor $C_f$.
The next step is the optimization of the $FoM$ with two parameters: the ratio of the tuner's reactance to transmission line characteristic impedance and the radius of the outer conductor of the tuner's resonator line $b_r$. The optimization simultaneously determines $b_r$ and the series capacitor $C_s$. Finally, the external Q of the tuner's port $Q_e$ is determined by the tuning range of the design, as based on measurements made on MESA cavities. 
Two MESA cavities were tested cryogenically. In these tests the power dissipated on the cavity walls was $P_\mathrm{diss}$ up to $\sim10$\,W each, at a stored energy $U$ of 15.3\,J \cite{TUSmesa}.  The microphonics detuning was measured as  $\Delta f_{\mu}$=$\pm$25\,Hz, determining the minimal tuner range. The thermal design of the tuner was set to allow up to 104 Hz to provide a safety margin.
The analytical analysis determines approximately all the necessary parameters of the tuner, enabling now a finite-element analysis of the precise values.

\section{Finite-Element Simulation}
The starting point is finite-element input file of the 1.3\,GHz 9-cell cavity model, which in our case was based on the dimensions reported in \cite{BAune}. The objectives are to adjust the dimensions to obtain the exact value of the resonant frequency $f_0$ and provide a port to couple the tuner to the cavity.  This cavity model has three ports, two of them are for a fundamental power coupler (FPC) and a pick up coupler (PU), with external quality factors $Q_{FPC}=1\cdot10^{9}$ and $Q_{PU}=1\cdot10^{11}$, respectively, both determined with their ports terminated by their characteristic impedance. A third port, with the same radii as the FPC port, was created for the FE-FRT. Initially, only the transmission line inner conductor of the tuner was connected to its port, and its penetration depth is varied until the desired value of $Q_e$, as determined with the analytical model, was reached, while its port was similarly terminated. It is worth noting that high-order mode damping ports and antennas were not included in this simulation. 
Following the analytical optimization and design, the structure of the tuner was created in a CST input file. The particular layout of the tuner follows the basic design concepts of the 400 MHz tuner provided in Figure 6 of  \cite{optim}. In particular, that design provides an elegant division of the tuner into two parts, with the same sapphire dielectric serving two purposes: The first element is a coupling port including a UHV vacuum sapphire window that can be assembled onto the cavity in a clean room. The second element is the tuner resonator, including the ferroelectric capacitor, which is isolated from the interior of the superconducting cavity by the sapphire window. The window thus also serves as the series capacitor $C_s$ which is part of the tuner's electrical circuit.
In the next step of the finite-element investigation, the tuner is simulated without the cavity, and the impedance is calculated at a CST port attached to the end of the quarter-wave resonator, where the transmission line leading to the cavity will be attached later on. 
 At this point, the inductance $L_f=\left|Z_{i}/{j\omega}\right|$ is adjusted, where $Z_i$ is the contribution to the inductive resistance in the tuner resonator \cite{optim}. Using the values of Table\,\ref{tab:table1} and the total parallel capacitance, given by Eq.\,\ref{eq:ctot}, determines the frequency of the tuner resonator $f_{\mathrm{res}}=1/\sqrt{4\pi^2CL_{\mathrm{f}}}=1.3$\,GHz, where the capacitance is given by

\begin{equation}
C=\frac{C_{\mathrm{f}}C_{\mathrm{s}}}{C_{\mathrm{f}}+C_{\mathrm{s}}}.\label{eq:ctot}
\end{equation}

\begin{figure}[!htbp]
	\centering
		\includegraphics[width=1.00\columnwidth]{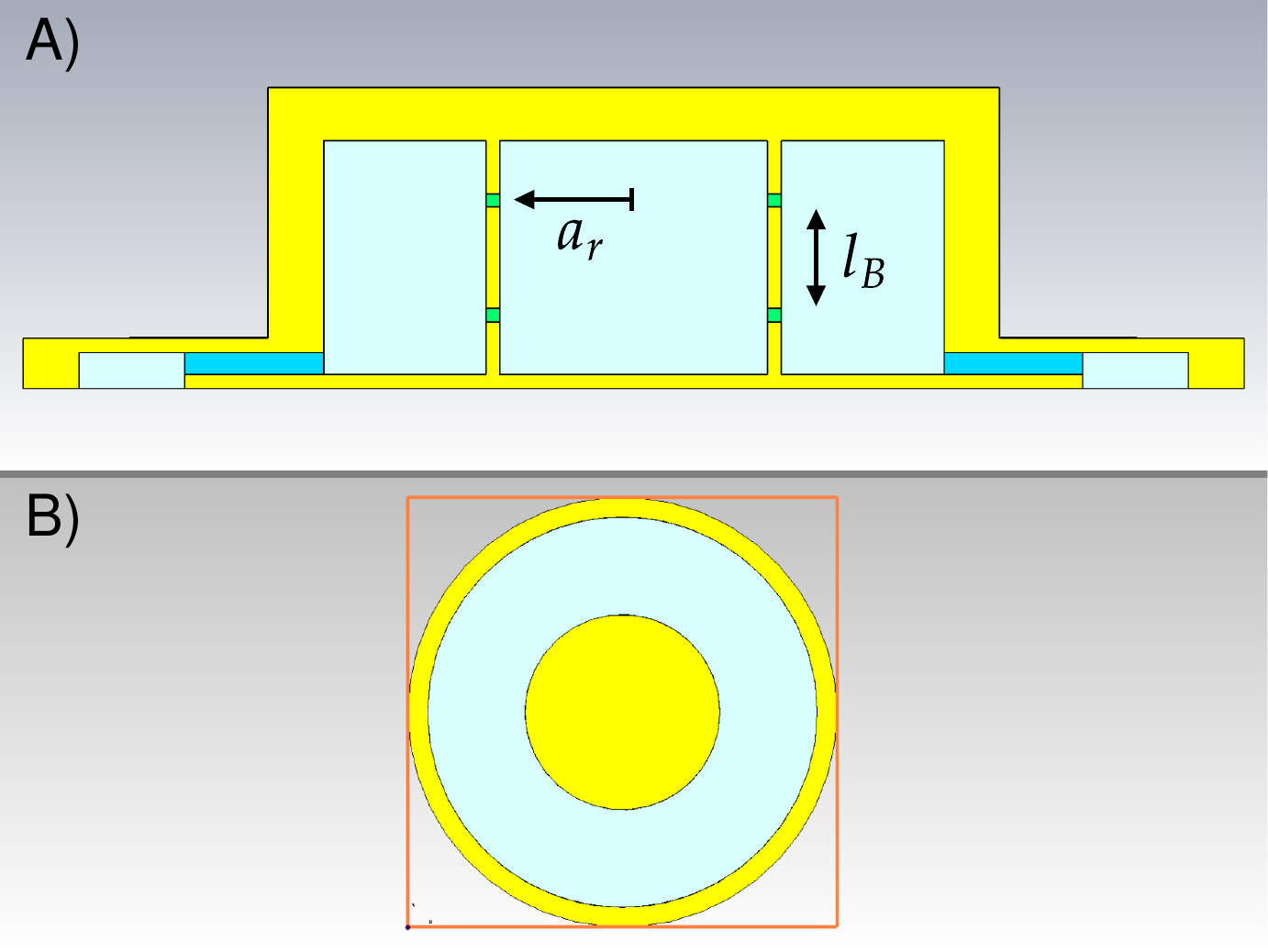}
	\caption{A) CST model of the turner design with a transmission line length of approximately 0\,mm. In this configuration, $a_r$ represents the inner radius of the ferroelectric wafers and $l_B$ the length of the central spacer. B) Bottom view of the tuner: the CST waveguide port (orange square) is placed at the termination of the coaxial line.}    	
	\label{fig:tuneres}
\end{figure}

The inductance is adjusted by changing $l_B$, the length of the spacer located between the two ferroelectric capacitors, indicated in Fig.\,\ref{fig:tuneres}.
This step is performed to readjust the geometry of the tuner resonator to match the frequency target, corresponding to the cavity's resonant frequency $f_0$. 

\begin{table}[!htbp]
\caption{Analytical tuner design parameters calculated using the Maple code. \label{tab:table1}}
\medskip{}
\centering{}%
\begin{tabular}{ccc}
\toprule 
Parameter & Analytical & Units\tabularnewline
\midrule
\midrule 
\smallskip
$f_{0}$ & 1.3 & GHz\tabularnewline
\smallskip
$U$ & 15.3 & J\tabularnewline
\smallskip
$\Delta f$ & 104 & Hz\tabularnewline
\smallskip
$A_{\mathrm{opt}}/g$ & 63.71 & mm\tabularnewline
\smallskip
$g$ & 0.5 & mm\tabularnewline 
\smallskip
$w$ & 0.5 & mm\tabularnewline
\smallskip
$a_{\mathrm{r}}$ & 4.82 & mm\tabularnewline
\smallskip
$b_{\mathrm{r}}$ & 5.32 & mm\tabularnewline
\smallskip
$Q_{\mathrm{e}}$ & 7.10 & $\cdot10^{7}$\tabularnewline
\smallskip
$\left|Z_{0}\right|$ & 41.59 & $\Omega$\tabularnewline
\smallskip 
 $a$ & 10 & mm\tabularnewline
\smallskip
$b$ & 20 & mm\tabularnewline
\bottomrule
\end{tabular}
\end{table}

\begin{figure}[!htbp]
	\centering
		\includegraphics[width=1.00\columnwidth]{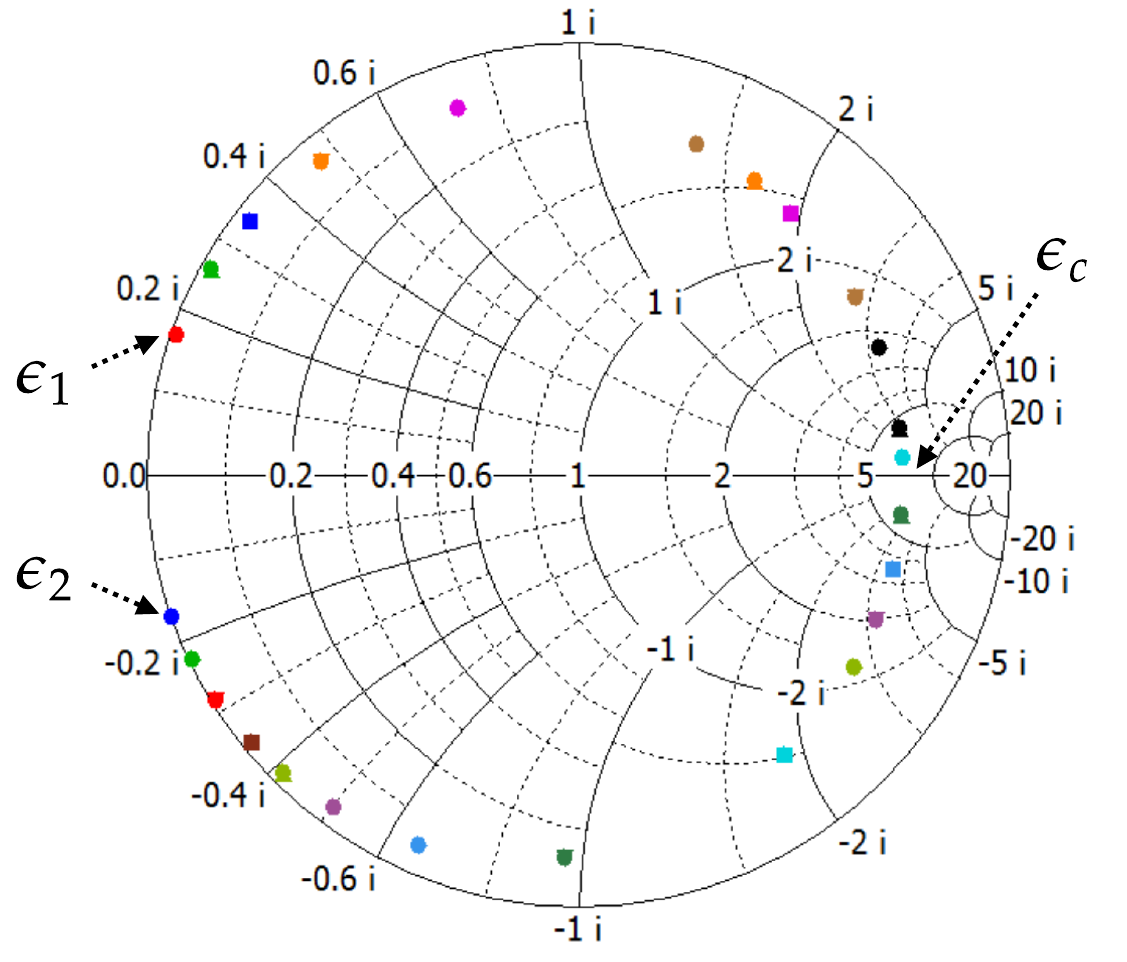}
	\caption{Symmetrized Smith chart for a transmission line length of $\sim0$\,mm. The impedances corresponding to state 1 and 2, and the central value, are indicated by $\epsilon_1$, $\epsilon_2$, and $\epsilon_c$, respectively.}    	
	\label{fig:scnotl}
\end{figure}

\begin{figure}[!htbp]
	\centering
		\includegraphics[width=1.00\columnwidth]{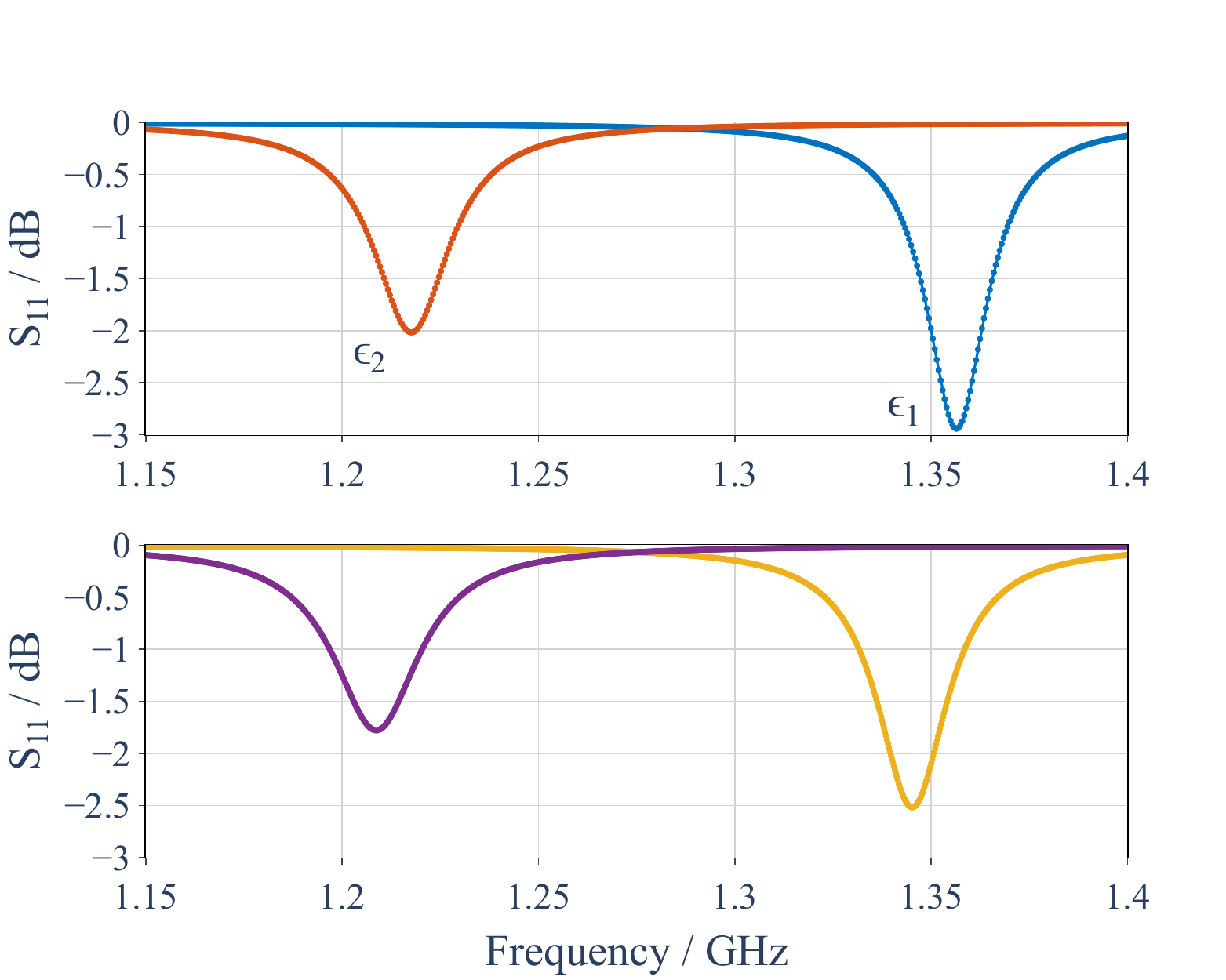}
	\caption{$S_{11}$ as a function of frequency for the two external values of the relative permittivity, $\epsilon_1$ and $\epsilon_2$, in the stand alone FRT simulation. The top subplot shows the analytical results, while the bottom subplot corresponds to the CST simulation data.}    	
	\label{fig:s11evsf}
\end{figure}

After determining the inductance, an $S_{11}$ Smith chart plot of the stand alone tuner is inspected as a function of the permittivity of the ferroelectric over its full range. A parameter sweep is performed changing the inner radius of the FE wafer $a_r$, which modifies the capacitance $C_f$, and the central spacer $l_B$ in the resonator, as indicated in Fig.\,\ref{fig:tuneres}. The optimal values in this parameter scan are those for which the imaginary part of the impedance $Z_L$ at the CST port at the two end states, the reactances $X_{1}=\Im\left[Z_{L}\left(\epsilon_{1}\right)\right]$ and $X_{2}=\Im\left[Z_{L}\left(\epsilon_{2}\right)\right]$, satisfy the condition $X_1=-X_2$. With this geometry, $Z_L$ as a function of the permittivity is already symmetric around the open position of the Smith chart as displayed in Fig.\,\ref{fig:scnotl}, making the tuning range $\Delta f$ symmetric around $f_0$. Note that the permittivity value for which $\Re\left[Z_{L}\left(\epsilon_{c}\right)\right]\rightarrow\infty$ and $\Im\left[Z_{L}\left(\epsilon_{c}\right)\right]=0$, defines the central relative permittivity $\epsilon_{c}$ given by $\approx\sqrt{\epsilon_{1}*\epsilon_{2}}$. The $S_{11}$ plot in dB calculated through the analytical, determined using Eq.\,\ref{eq:refc} \cite{pozar} with the impedance of the transmission line $Z_0$, and simulation models for $\epsilon_{1}$ and $\epsilon_{2}$ are shown in Fig.\,\ref{fig:s11evsf}.

\begin{equation}
S_{11}[\mathrm{dB}]=20\cdot\mathrm{log}_{10}\left(\left|\frac{Z_{L}-Z_{0}}{Z_{L}+Z_{0}}\right|\right)\label{eq:refc}
\end{equation}

The finite-element electromagnetic simulation provides adjustment to the exact dimensions of the tuner, which may vary slightly from the analytic values.  

\begin{figure}[!htbp]
	\centering
		\includegraphics[width=1.00\columnwidth]{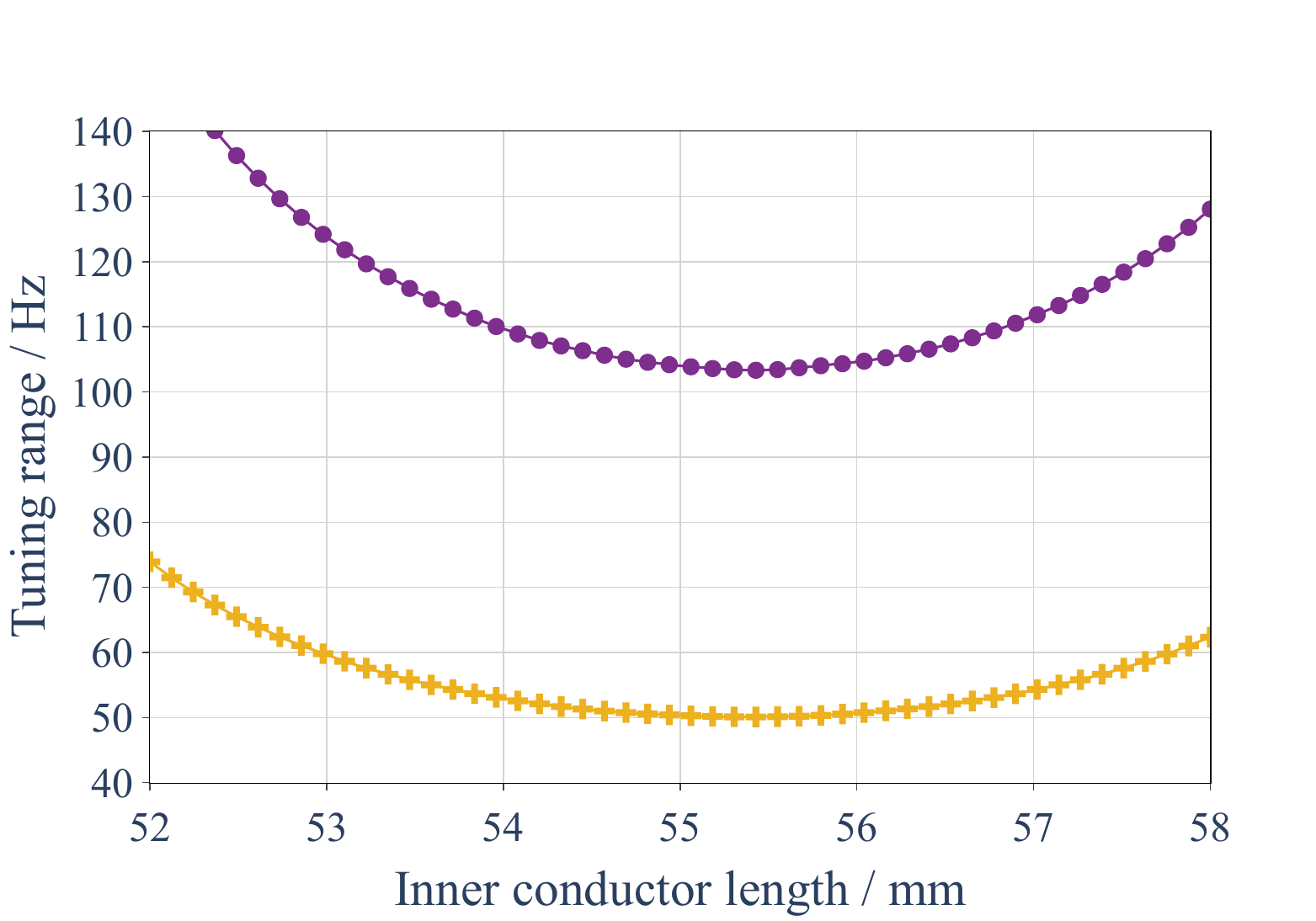}
	\caption{Tuning range versus transmission line length for two tuner designs, showing minimum values of 50\,Hz (yellow plus signs) and 104\,Hz (purple circles).}    	
	\label{fig:trvslng}
\end{figure}

\begin{figure}[!htbp]
	\centering
		\includegraphics[width=1.00\columnwidth]{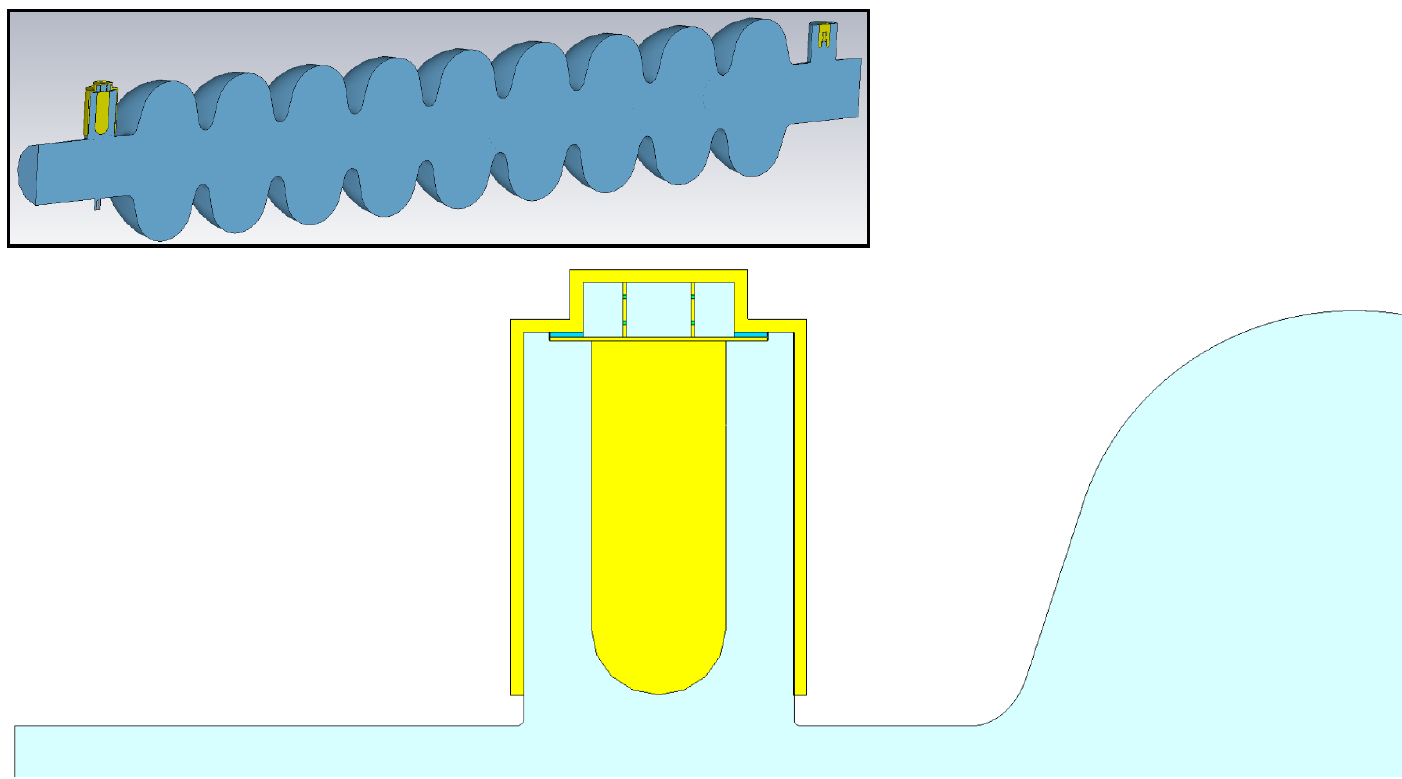}
	\caption{CST model illustrating the tuner installed to a 1.3\,GHz 9-cell SRF cavity, including the fundamental power and pick-up couplers.}    	
	\label{fig:cavfrt}
\end{figure}

The analytic design includes a quarter-wave transmission line between the tuner's resonator and the cavity's port. At 1.3\,GHz, a quarter-wave is 57.7\,mm long. The coupler forms this quarter-wave transmission line. However, the exact length of the probe depends on the electric field distribution at the tip of the probe, and is not easily amenable to an analytical calculation. Therefore, this is done following a procedure \cite{optim} where the tuning range is plotted as a function of the length of the transmission line, as shown in Figure \ref{fig:trvslng}. The length of the line is determined to be at the minimal value of the tuning range. 

This procedure is carried out when the tuner is incorporated into the 1.3\,GHz 9-cell cavity model, see Fig.\,\ref{fig:cavfrt}. At first, an inner conductor almost as long as a quarter wavelength is considered. Then, its length is adjusted without changing the distance of the tip of the probe from the cavity's center line, thus not affecting the $Q_e$ of the coupling port. In our case, the length of the inner conductor was $\sim55.5$\,mm instead of the original estimate of 57.7\,mm. At this point, the geometry of the tuner is completely determined and other variables are obtained such as the frequency differences $\Delta f_1=f\left(\epsilon_1\right)-f_0$ and $\Delta f_2=f\left(\epsilon_2\right)-f_0$, and the Q-factor of the tuner $Q_{FRT}\left(\epsilon\right)$ as a function of the relative permittivity. Notice that since $f_0$ is between $f\left(\epsilon_1\right)$ and $f\left(\epsilon_2\right)$, this leads to $\Delta f_1>0$ and $\Delta f_2<0$. The $Q_{FRT}\left(\epsilon\right)$ considers the surface losses on the tuner Cu structure and the dielectric losses on the ferroelectric wafers, and it is deduced from the loaded cavity quality factor $Q_L$, given by the following equation, Eq.\,\ref{eq:qfrte},

\begin{equation}
\frac{1}{Q_{L}}=\frac{1}{Q_{0}}+\frac{1}{Q_{FRT}\left(\epsilon\right)}+\frac{1}{Q_{FPC}}+\frac{1}{Q_{PU}},\label{eq:qfrte}
\end{equation}

where $Q_0$ is the unloaded cavity quality factor, accounting for the surface losses in its walls. The tuner quality factors at states 1 and 2, $Q_{FRT,1}=Q_{FRT}\left(\epsilon_1\right)$ and $Q_{FRT,2}=Q_{FRT}\left(\epsilon_2\right)$ respectively, are important as they, along with the tuning range, are used to calculate the $FoM$ as shown in Eq.\,\ref{eq:fom}.

\begin{equation}
FoM=\frac{\Delta f}{f_{0}}\overline{Q}_{\mathrm{FRT}}\label{eq:fom}
\end{equation}

Here, the average FRT quality factor $\overline{Q}_{\mathrm{FRT}}$ is defined by Eq.\,\ref{eq:Qavg}.

\begin{equation}
\overline{Q}_{\mathrm{FRT}}=\sqrt{Q_{\mathrm{FRT},1}*Q_{\mathrm{FRT},2}}\label{eq:Qavg}
\end{equation}

A summary of the tuner parameters optimized for a tuning range $\Delta f=104$\,Hz are displayed in Table\,\ref{tab:table2}. The analytical lumped element circuit analysis shows good agreement with the values determined by CST. It is worth mentioning that the capacitances $C_f$ and $C_s$ were individually calculated with a simplified geometry in an electrostatic simulation.
It is important to realize that the values of the $FoM$ and $Q_{FRT}$ depend on two elements. One is the performance of the circuit.  The other is the quality of the ferroelectric material, which includes the the loss tangent of the ferroelectric material and the tunability of the ferroelectric material. Thus, the $FoM$ of the complete tuner is proportional to the intrinsic $FoM(FE)$ defined as
\begin{equation}
FoM(FE)=\frac{ \epsilon_2 -\epsilon_1}{2 \delta \epsilon_c}.\label{eq:fom(FE)}
\end{equation}

Therefore, in comparing the $FoM$ values of two tuners, one must consider using the same material performance, that is the $FoM(FE)$ value. In this work the conservative value used is $FoM(FE)=62.12$, where $\epsilon_c\approx111.78$.

\begin{table}[!htbp]
\caption{Comparison of tuner design parameters obtained from the analytical model and CST simulations. \label{tab:table2}}
\medskip{}
\centering{}%
\begin{tabular}{cccc}
\toprule 
Parameter & Analytic & CST & Units\tabularnewline
\midrule
\midrule 
\smallskip
$\Delta f$ & 104 & 104.01 & Hz\tabularnewline
\smallskip
$C_{\mathrm{f}}$ & 15.74 & 16.07 & pF\tabularnewline
\smallskip 
$C_{\mathrm{s}}$ & 45.39 & 45.40 & pF\tabularnewline
\smallskip 
$l_{\mathrm{r}}$ & 8.84 & 8.54 & mm\tabularnewline
\smallskip
$Q_{\mathrm{FRT},1}$ & 2.54 & 2.67 & $\cdot10^{8}$\tabularnewline
\smallskip
$Q_{\mathrm{FRT},2}$ & 1.49 & 1.17 & $\cdot10^{9}$\tabularnewline
\smallskip
FoM & 49 & 44.19 & -\tabularnewline
\smallskip
$\left|\Delta f_{1}\right|$ & 52 & 49.98 & Hz\tabularnewline
\smallskip 
$\left|\Delta f_{2}\right|$ & 52 & 56.89 & Hz\tabularnewline
\smallskip 
$Q_{\mathrm{e}}$ & 7.10 & 7.70 & $\cdot10^{7}$\tabularnewline
\bottomrule
\end{tabular}
\end{table}

\begin{figure}[!htbp]
	\centering
		\includegraphics[width=1.00\columnwidth]{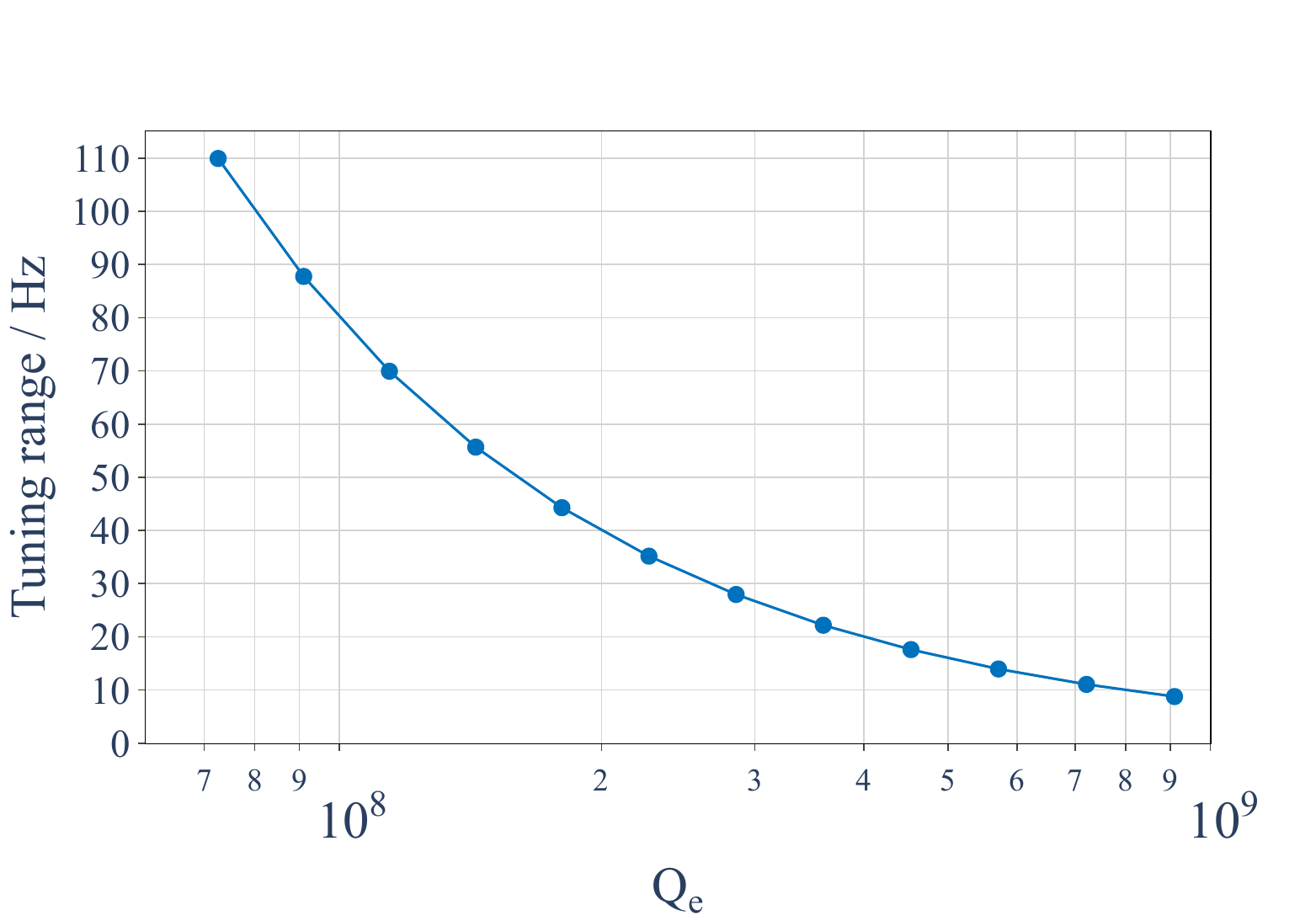}
	\caption{Tuning range as a function of the tuner port external quality factor $Q_e$.}    	
	\label{fig:qevstr}
\end{figure}

Given that the initial tuning range was chosen approximately conservatively as two times the detuning $\Delta f_\mu$, which is comparable to the cavity bandwidth with an $Q_{FPC}=1.25\cdot10^7$, smaller tuning ranges were calculated as well for comparison. The Q-factors and $FoM$ for tuning ranges between 5 and 100 are shown in Table\,\ref{tab:table3}. This is relevant as other facilities currently employing 1.3\,GHz 9-cell cavities have reported various detuning values. For example, SLAC LCLS-II cavities are affected by $<$30\,Hz detuning \cite{lclsii}, while the European XFEL cavities show up to 6\,Hz with piezoelectric tuners \cite{euxfel}. The process here follows the readjustment of the inner conductor length at the minimum of the tuning range as described above in Fig.\,\ref{fig:trvslng}. Then, with this length fixed, the penetration of the inner conductor tip was varied to modify the $Q_e$. The dependence of the tuning range on $Q_{e}$ are shown in Fig.\,\ref{fig:qevstr}. Note that the $f_0$ was recalculated for every data point due to the tuner port changing in length by a rage of approximately 15\,mm. Conversely, the $FoM$ remains constant since it does not depend on the strength of the tuner’s coupling as shown in Table\,\ref{tab:table3}.

\begin{figure}[!htbp]
	\centering
		\includegraphics[width=1.00\columnwidth]{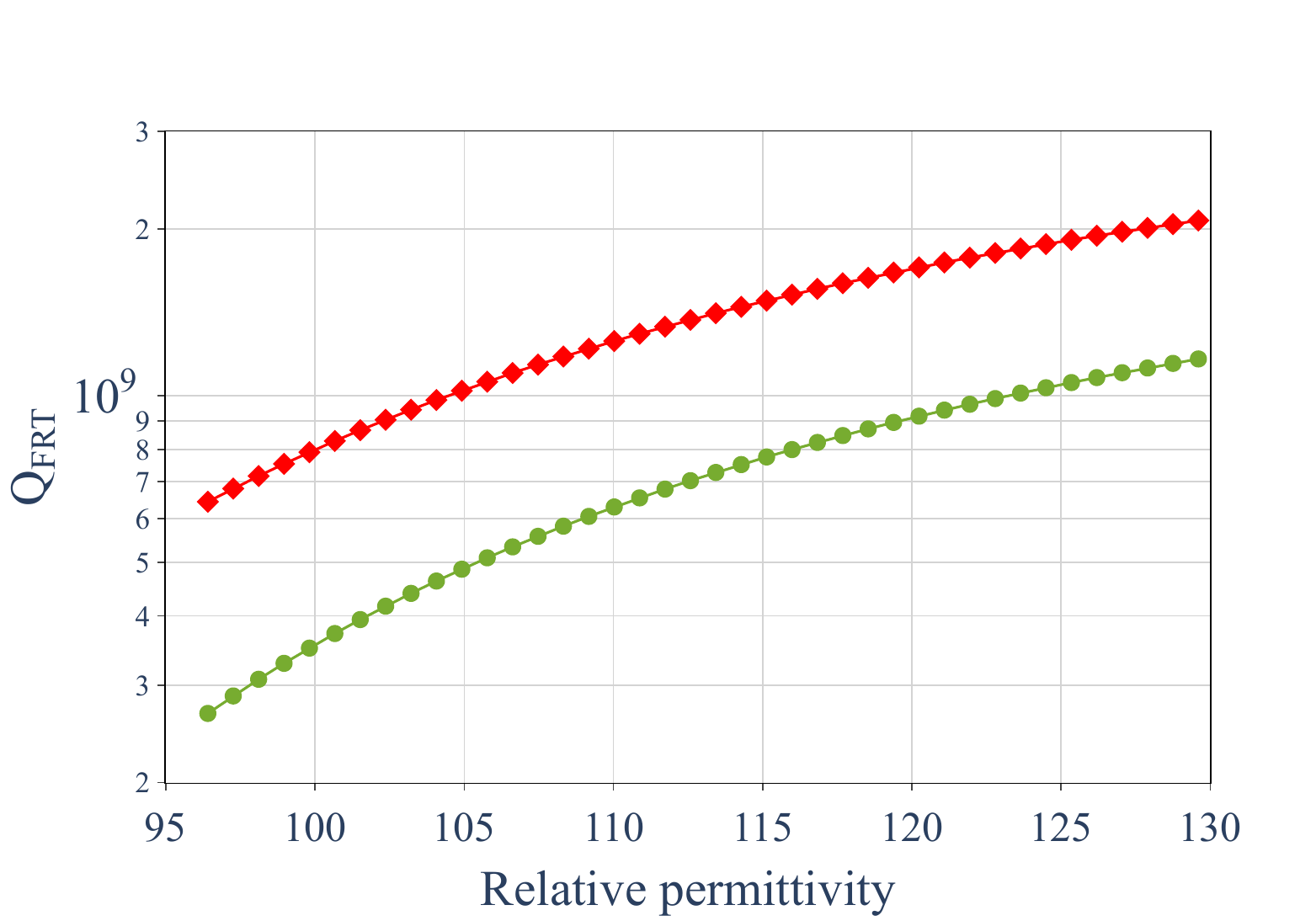}
	\caption{Tuner internal quality factor $Q_{FRT}$ versus relative permittivity for the two designs with minimum tuning ranges of 50\,Hz (red diamonds) and 104\,Hz (green circles).}    	
	\label{fig:qfrtvsep}
\end{figure}

One must also consider how well does the $FoM$ represent the performance of the tuner. The usual definition of the $FoM$ describes the performance of the tuner as a single, average number, as seen for example in Eq.\,\ref{eq:fom}. For a cavity with a tuner, where the RF power amplifier  requirement is determined  by the highest power dissipation in the tuner, this average indicator is not appropriate. Looking at Figure\,\ref{fig:qfrtvsep}, we see $Q_{FRT}$ plotted as a function of the permittivity for the cases of the 50\,Hz and 104\,Hz tuning range. This information is necessary to specify the power rating of the RF amplifier driving the cavity. The amplifier must be able to address the requirement at the lowest value of $Q_{FRT}$, otherwise the cavity voltage would not be maintained at all values of the detuning. It is obvious that this condition is specified by the value of $Q_{FRT}$ at the lowest permittivity.

\begin{table}[!htbp]
\caption{Tuner design parameters corresponding to tuning ranges of 10, 25, 50, 75, and 100\,Hz. \label{tab:table3}}
\medskip{}
\centering{}%
\begin{tabular}{ccccc}
\toprule 
$\Delta f$/Hz & $Q_{\mathrm{e}}$/$10^{7}$ & $Q_{\mathrm{FRT,1}}$/$10^{8}$ &  $Q_{\mathrm{FRT,2}}$/$10^{9}$ & FoM\tabularnewline
\midrule
\midrule 
10.01 & 79.61 & 27.78 & 11.83 & 44.13\tabularnewline
25.01 & 31.88 & 11.08 & 4.76 & 44.17\tabularnewline
50.02 & 15.95 & 5.51 & 2.39 & 44.22\tabularnewline
75.03 & 10.64 & 3.65 & 1.60 & 44.20\tabularnewline
100.00 & 8.01 & 2.73 & 1.21 & 44.19\tabularnewline
\bottomrule
\end{tabular}
\end{table}


For an energy-recovery accelerator with negligible beam loading such as MESA, the forward RF power needed to maintain the accelerating voltage $V_c=12.97$\,MV is given by Eq.\,\ref{eq:Prf} \cite{erlPrf},

\begin{equation}
P_{RF}=\frac{\left(1+\beta\right)^{2}}{4\beta Q_{0}}\frac{V_{c}^{2}}{\nicefrac{R}{Q}}\left[1+\left(\frac{2Q_{0}}{1+\beta}\frac{\Delta f_{\mu}}{f_{0}}\right)^{2}\right],\label{eq:Prf}
\end{equation}

where $R/Q=1030\,\Omega$ and $\Delta f_\mu$ is cavity detuning. To minimize the forward RF power, $Q_{FPC}$ must be matched to the heaviest load inside the cavity, which primarily corresponds to the effective Q-factor resulting from the combined dissipated power in the walls and in the FRT. Therefore, $Q_0$ used in Eq.\,\ref{eq:Prf} should account for these power losses, making it dependent on the relative permittivity. Accordingly, the coupling $\beta\left(\epsilon\right)=Q_0\left(\epsilon\right)/Q_{FPC}$ is determined by the lowest $Q_{FRT}$. Therefore, to quantify the coupling strength $\beta\left(\epsilon\right)$, and thus the forward power $P_{RF}$, $Q_{FRT}$ as a function of the permittivity was determined, see Fig.\,\ref{fig:qfrtvsep}. As previously discussed, the FPC is critically coupled ($\beta=1$) to $Q_{FRT,1}$ and becomes increasingly over-coupled until $Q_{FRT,2}$ is reached. It is important to remember that at state 1 or $\epsilon_1$, the maximum bias voltage is applied across the ferroelectric wafer, aligning its electric dipoles, which reduces the capacitance $C_f$ and increases the dielectric's effective loss. Conversely, in state 2 or $\epsilon_2$, with no voltage applied, dipoles alignment relaxes, resulting in an increased $C_f$ and reduced dielectric dissipation. Additionally, the differences in $Q_{FRT}$ values for the two tuning ranges shown in Fig.\,\ref{fig:qfrtvsep} are due to the fact that, in the 50\,Hz case, the inner conductor tip is positioned approximately 3\,mm farther from the cavity center compared to the 104\,Hz tuning range design.

The maximal forward RF power $P_{RF}^{FRT}$ required to drive the cavity with the FRT for a tuning range of 50\,Hz is 267\,W.  In contrast, for a cavity without an FRT, with optimal coupling strength given by

\begin{equation}
\beta_{opt}=\sqrt{1+\left(2Q_{0}\frac{\Delta f_{\mu}}{f_{0}}\right)^{2}},\label{eq:betanofrt}
\end{equation}
which is derived by optimizing Eq.\,\ref{eq:Prf}, the minimum required power $P_{RF}$ is 3150\,W. The ratio of these powers $P_{RF}/P_{RF}^{FRT}$ gives a value of approximately 12. 

\section{Conclusions}
This work presents the design of an ferroelectric fast-reactive tuner for a 1.3\,GHz 9-cell TESLA type cavity of the Mainz Superconducting Energy-Recovering  Linac. The design to compensate the measured acoustic detuning of $\pm$25\,Hz is through over-coupling the RF power. In order to reduce the RF power needed for this correction, a ferroelectric reactive tuner is being investigated. Considering the cavity stored energy, its resonant frequency and the required FRT tuning range, this design was developed following the procedure described in \cite{optim}. An analytic optimized design was obtained using a lumped-element circuit model. Then this design was implemented in finite-element simulations using CST Microwave Studio. Excellent agreement for the tuner's parameters and performance is observed between the two methods.  The initial choice for the tuning range was conservatively set to 104\,Hz. This range choice corresponded to that of the original bandwidth designed for the MESA cavities. However, following frequency noise measurements, the tuning range was adjusted to 50\,Hz. However, the conservative tuner optimization for 104\,Hz tuning range  is quite effective for operation at the 50\,Hz range providing a good Figure of Merit. In the simulations presented here, it is shown that while a cavity without an FRT, and with an optimized coupling strength of the FPC, requires an RF power of 3150\,kW, adding an FRT to a dedicated port in the cavity lowers the power to under 0.3\,kW. 

The  results reported here demonstrate a good performance of the tuner design for a 50\,Hz tuning range. FE-FRTs are powerful tools developed to mitigate the negative effects of microphonics in SRF cavities and reducing the wall-plug power consumption required for their operation. Ongoing FRT research for such application aims to further enhance their energy-saving property.

\section{Acknowledgements}
This work was supported by the BMBF under the research grants 05H24UM1.  
   
\bibliographystyle{naturemag}

\end{document}